\documentclass[
aip,
amsmath,amssymb,
reprint,%
]{revtex4-1}
\usepackage[colorlinks=true,
linkcolor=red,
urlcolor=black,
citecolor=blue]{hyperref}
\usepackage{graphicx}
\usepackage{epstopdf}
\usepackage{dcolumn}
\usepackage{bm}

\usepackage{color}
\usepackage{soul}
\definecolor{darkgreen}{rgb}{0.1,.6,.1}
\definecolor{greenblue}{rgb}{0.0,.1,.4}

\usepackage[normalem]{ulem}

\begin{document}


\title {Transition to hyperchaos: Sudden expansion of attractor and intermittent large-amplitude events in dynamical systems}
	\author{S. Leo Kingston}
	\thanks{Corresponding author:~kingston.cnld@gmail.com}
\affiliation{Division of Dynamics, Lodz University of Technology, 90-924 Lodz, Poland}	

\author{Tomasz Kapitaniak}
 \affiliation{Division of Dynamics, Lodz University of Technology, 90-924 Lodz, Poland}
\author{Syamal K. Dana}
\affiliation{Division of Dynamics, Lodz University of Technology, 90-924 Lodz, Poland}
\affiliation{Department of Mathematics, National Institute of Technology, Durgapur 713209, India}

\date{\today}




\begin{abstract} Hyperchaos is distinguished from chaos by the presence of at least two positive Lyapunov exponents instead of just one in dynamical systems. 
A general scenario is presented here that shows emergence of hyperchaos  with a sudden large expansion of the attractor of continuous dynamical systems at a critical parameter when the temporal dynamics shows intermittent large-amplitude spiking or bursting events. The distribution of local maxima of the temporal dynamics is non-Gaussian with a tail, confirming a rare occurrence of the large-amplitude events. 
We  exemplify our results on the sudden  emergence of hyperchaos in three paradigmatic models, namely, a coupled Hindmarsh-Rose model, three coupled Duffing oscillators, and a hyperchaotic model. 
\end{abstract}

\maketitle
\begin{quotation}
Hyperchaos has been observed in many dynamical systems  (discrete or continuous), which is recognized by the presence of at least two positive Lyapunov exponents. Different dynamical processes are involved in the origin of  hyperchaos from periodic, quasiperiodic or chaotic orbits in response to a system parameter. Hyperchaos appears with a discontinuous large expansion of the attractor of a system at a critical parameter. One after another Lyapunov exponents usually transits to positive values against a parameter change when hyperchaos appears from chaos. In many other systems, hyperchaos appears from a periodic or quasiperiodic orbit when two Lyapunov exponents simultaneously go to a positive value.  In particular, the temporal dynamics carries a signature of intermittent large-amplitude events often called as extreme events. We provide here numerical evidence of  this scenario in three  paradigmatic continuous dynamical systems of a higher dimension.
\end{quotation}

\section{Introduction}

\par   Poincar\'e \cite{poincare1967new} reported in 1890 that the trajectory of three celestial bodies is sensitive to their initial conditions. In a deterministic dynamical model, Lorenz \cite{lorenz1963deterministic} confirmed the sensitive dependence of the trajectory of the system in 1963 which was later defined as chaos \cite{li2004period}. It  has been characterized by the presence of one positive Lyapunov exponent in a three-dimensional (3D) system that is responsible for directional stretching of the system's trajectory \cite{alligood1996chaos} and one negative exponent that leads to folding of the trajectory, thereby making a trajectory remained bounded within a volume of state space of the system. There exists one more Lyapunov exponent of zero value showing no folding or stretching of the phase of the system. Later, in a four-dimensional system \cite{rossler1979equation}, R\"ossler identified the existence of two positive Lyapunov exponents besides one negative and one zero Lyapunov exponents and  defined the dynamics as hyperchaos. 
\par The appearance of hyperchaos in high-dimensional
 systems has been well documented, by this time, in many systems \cite{ franco2020crisis, miranda2013lagrangian, kapitaniak1993transition, perlikowski2010routes, kengne2015coexistence, tamasevicuius1997hyperchaotic, colet1994controlling, stankevich2018chaos, stankevich2019chaos,kashyap2020hyperchaos,stankevich2021chaos,garashchuk2019hyperchaos, stankevich2020scenarios}, and the mechanisms have also been explored with identification of  the distinct dynamical processes involved in the origin of hyperchaos.  The occurrence of hyperchaos is connected to a secondary Neimark-Sacker bifurcation followed by torus breakdown in a radio-frequency generator \cite{stankevich2019chaos} and boundary crisis in a four-dimensional R\"ossler system \cite{stankevich2020scenarios}. In a ring of three synthetic genetic oscillators with quorum-sensing coupling, hyperchaos arises due to merging of unstable  cyclic rotating waves with a chaotic attractor \cite{stankevich2021chaos}. The existence of hyperchaos has been reported in  two coupled gas bubble models based on the emergence of a homoclinic chaotic attractor containing a saddle-focus periodic orbit with its two-dimensional unstable manifold \cite {garashchuk2019hyperchaos}. Hyperchaos is also found in the nonlinear chimney model, and it is used to understand the complex swaying dynamics of trees \cite{kashyap2020hyperchaos}. A sudden expansion of the attractors of the systems was always noticed during the transition to hyperchaos, and two Lyapunov exponents become positive one after another in response to a change of a parameter. 
In recent times, hyperchaotic polarized light has been identified in a free-running laser diode that elucidates highly intricate and irregular laser emission with a variety of statistical distributions of the light intensity  \cite {bonatto2018hyperchaotic}. Once again the transition to hyperchaos shows two Lyapunov exponents becoming positive one after another with a variation of the injection current.   
\par Very recently, hyperchaos was reported in a semiconductor superlattice  \cite{mompo2021designing} that is smaller in size and equally faster compared to optical devices. It was clearly shown there that a sudden transition to a large-amplitude oscillation appears at a critical parameter via Pomeau-Manneville intermittency. The temporal dynamics shows occasional 
large-amplitude events; however, they did not emphasize on the specific parameter region of the appearance of hyperchaos and on another parameter region where two Lyapunov exponents transit simultaneously to a positive value. Moreover, they have not discussed about the statistical properties of the dynamics. On the other hand, in a Zeeman laser model, we reported \cite{kingston2022transition} origin of hyperchaos that follows three different dynamical routes, Pomeau-Manneville intermittency\cite{pomeau1980intermittent,kingston2017extreme} from a periodic orbit, quasiperiodic intermittency from a quasiperiodic orbit \cite{redondo1997intermittent,kingston2021instabilities} and crisis-induced intermittency from chaos \cite{grebogi1987critical,mishra2020routes}. 
For all the three specific routes to hyperchaos, a sudden large expansion of a regular (periodic or quasiperiodic) or chaotic attractor was found at a critical parameter.  The origin of hyperchaos was detected when two Lyapunov exponents  simultaneously switch to a positive value and when it appears from a periodic orbit via Pomeau-Manneville intermittency  and a quasiperiodic states via quasipeirodic intermittency. On the other hand, when hyperchaos appears from chaos via crisis-induced intermittency, the largest Lyapunov exponent is already positive and the second one goes to a positive value at a critical parameter when the large expansion of the attractor is clearly seen. The discontinuous large expansion of the attractor during the origin of  hyperchaos may be hysteresis-free or not, which depends upon the system under consideration. The temporal dynamics of hyperchaos, in particular, has a manifestation of rare and recurrent large intensity pulses while the amplitude of the laser remains confined to a bounded limit for most of the time. The distribution of the intensity pulses follows a non-Gaussian statistics with a tail. 
\par The discontinuous transition to hyperchaos in response to a system parameter and the recurrent large intensity pulses in the temporal dynamics  with a non-Gaussian statistics  are not restricted to the Zeeman laser, and it is rather a common  property of many other systems as we illustrate here from our numerical studies on  three paradigmatic models, a coupled Hindmarsh-Rose neuron model, three coupled Duffing oscillators, and a hyperchaotic model. 

 \section{Coupled Hindmarsh-Rose Neuron model}
We start our first numerical experiment on the emergence of hyperchaos in a coupled Hindmarsh-Rose neuron model  \cite{mishra2018dragon, mishra2020routes, chowdhury2022extreme} that is triggered by Pomeau-Manneville intermittency. The coupled neuron model is described as 
\begin{eqnarray}
\dot{x_{i}} &=& y_{i} + bx_{i}^{2} - ax_{i}^{3} - z_{i} + I - k_{i}(x_{i} - v_{s})\Gamma(x_{j})\\ \nonumber
\dot{y_{i}} &=& c - dx_{i}^{2} - y_{i}\\ \nonumber
\dot{z_{i}} &=& r[s(x_{i} - x_{R}) - z_{i}] \nonumber
\label{HR_eqn}
\end{eqnarray}
where $i,j=1,2$ ($i\neq j$) represents two oscillators. The coupling function is described by chemical synaptic interactions between the neurons as expressed by a sigmoidal function $\Gamma(x) = \frac{1}{1 + \exp^{-\lambda(x - \Theta)}}$. Parameter values of the system are $a$ = 1.0, $b$ = 3.0, $c$ = 1.0, $d$ = 5.0, and $x_{R}$ = -1.6. The isolated neurons show periodic bursting. The parameters of the coupling function are chosen as $v_s$ = 2.0, $\lambda$ = 10.0, $\Theta$ = -0.25, and the coupling strength $k_1$=$k_2$ is considered positive. 
\begin{figure}
	\begin{center}
		\includegraphics[width=0.85\columnwidth]{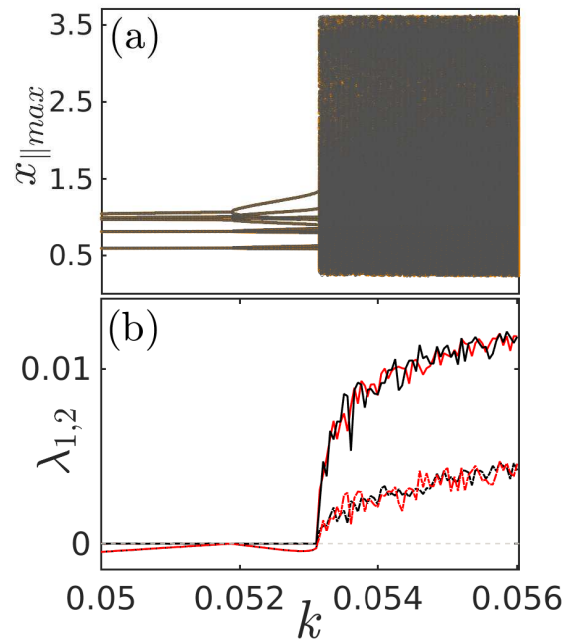}
	\end{center}
	\caption{Hyperchaos in a coupled Hindmarsh-Rose neuron model. (a) Bifurcation diagram for forward (yellow) and backward (gray) integrations  and two largest Lyapunov exponents (black and red lines) for the range of $k$ (0.050, 0.056). At  a critical parameter $k$ = 0.0532, (a) a sudden large expansion in $x_{\parallel max}$ is reflected in the bifurcation diagram  and (b) two Largest Lyapunov exponents become positive exemplifying the origin of hyperchaos. Red and black lines denote forward and backward integration indicating a hysteresis-free transition. Lyapunov exponents are estimated for $1.0\times10^7$ iteration time and after discarding $5.0\times10^6$ transients.}	\label{HR_PMI_BILYA} 
\end{figure}
\par
\begin{figure}
	\includegraphics[width=0.8\columnwidth]{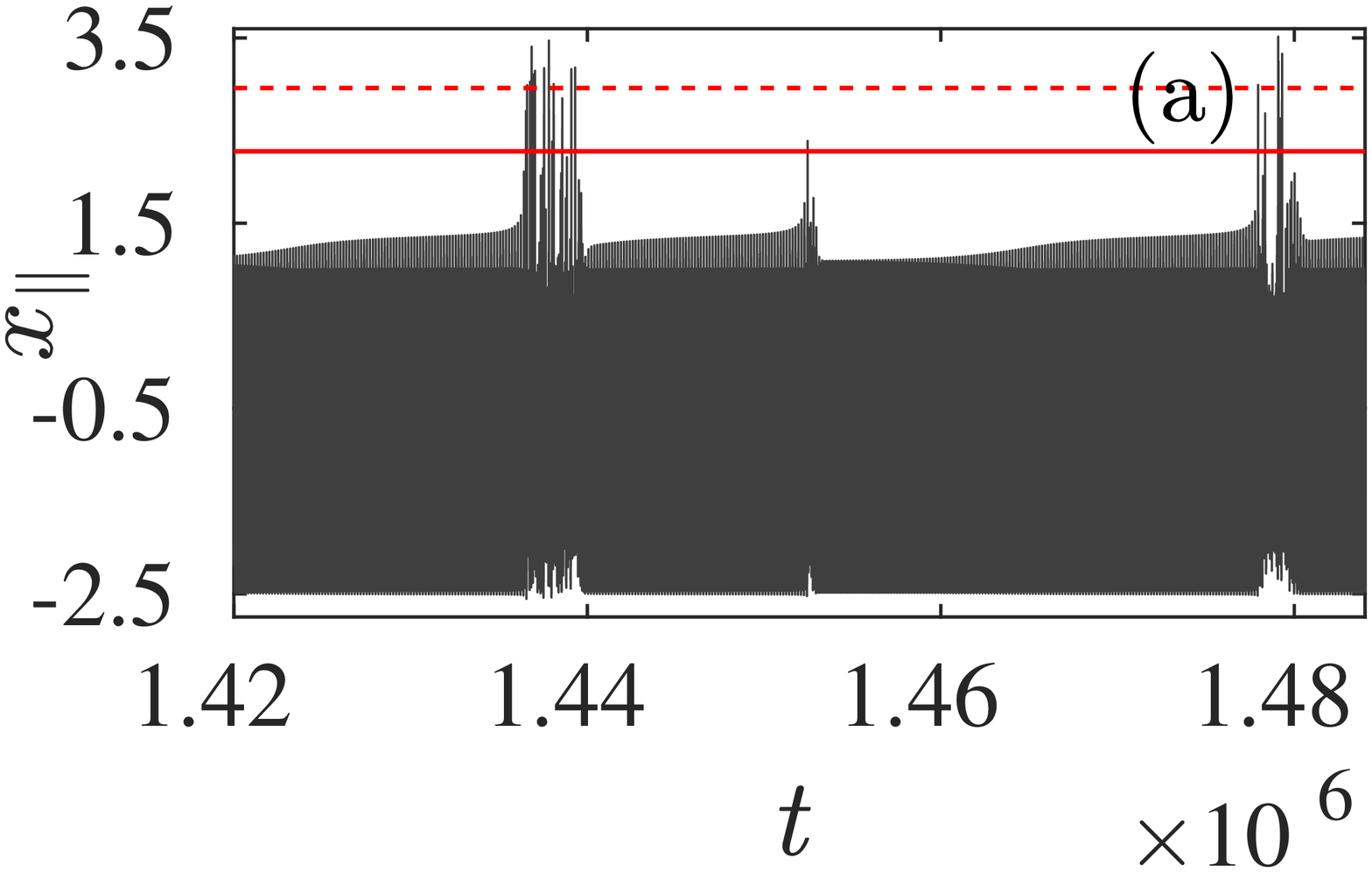}\\
	\includegraphics[height=4cm, width=5cm]{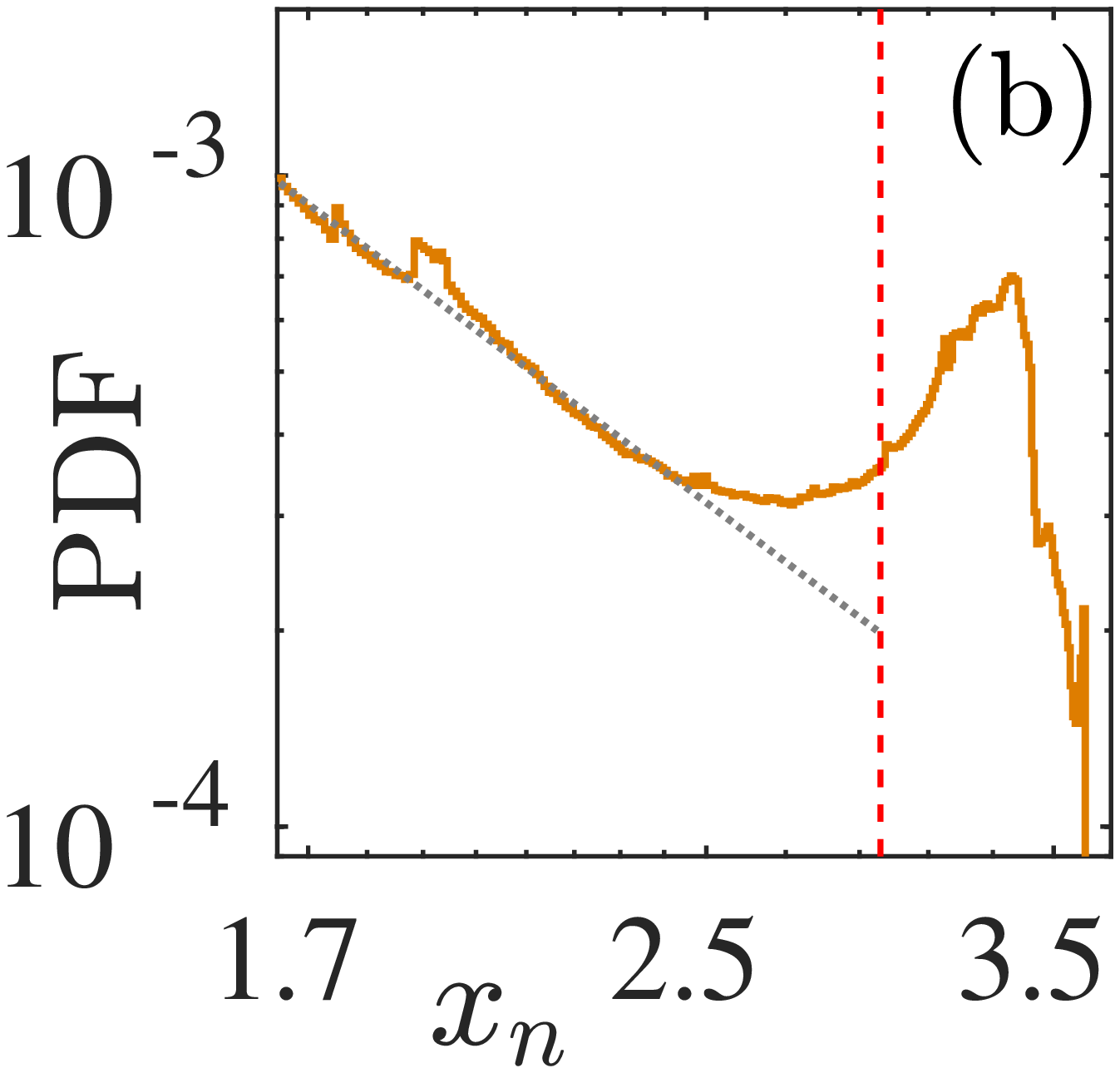}
	\caption{(a) Time evolution of hyperchaos in a coupled Hindmarsh-Rose model for $k$ = 0.0532. (b) Probability density distribution (PDF) of all the peaks in a time series in a log-log scale shows a dragon-king like bump. A  dashed horizontal line (red line) in (a)  and a vertical line (b) represent the significant height, $h_{s_1}$=$\langle x_{\parallel max}\rangle$ + $6 \sigma$. A solid horizontal line in (a)  signifies $h_{s}$=$\langle x_{\parallel max}\rangle$ + $4 \sigma$.}%
	\label{HR:TIME_PDF}
\end{figure}

 In a coupled state, 
this neuron model shows periodic bursting for a range of $k \in (0.050, 0.0531)$ values and our choice of system parameters. At a critical $k$ value, rare and recurrent large intermittent chaotic bursting appears. For a visual description of the transition, we draw a  bifurcation diagram of 
$x_{\parallel max}$ against the coupling parameter $k$ in Fig.~\ref{HR_PMI_BILYA}(a), where $x_\parallel$ = $x_1 + x_2$. The coupled system shows higher order periodicity  for a range of $k$ values, and then, a sudden discontinuous large expansion of $x_{\parallel max}$ occurs at a critical $k=0.0532$. Two largest Lyapunov exponents $\lambda_{1,2}$ of the coupled system become positive simultaneously as depicted in Fig.~\ref{HR_PMI_BILYA}(b). The  Lyapunov exponents of the system are calculated using a perturbation method \cite{kingston2021instabilities,balcerzak2018fastest}. The total integration time of the simulation is  $1.0\times 10^{7}$ and we remove a transient time of  $5.0\times 10^{6}$. 
Hyperchaos suddenly appears with a discontinuous transition from a periodic state against the variation of $k$. This transition shows no hysteresis as we checked with both forward and backward integrations (red and black lines). The system reveals almost a periodic motion in the time evolution of $x_{\parallel}$ but interrupted by occasional large-amplitude bursting as shown in Fig.~\ref{HR:TIME_PDF}(a). The typical signature of Pomeau-Manneville intermittency appears at a critical $k$, and it signifies the origin of hyperchaos that coincides with the emergence of occasional bursting. The remarkable new feature of intermittent bursting here is the height of the spikes within the bursts, which are larger than a significant height $h_{s}=\langle x_{\parallel max} \rangle+4\sigma$ (red horizontal line), where $\langle x_{\parallel max} \rangle$ is the mean of all local peaks in a temporal dynamics and $\sigma$ denotes the standard deviation.
A higher  significant height $h_{s_1}=\langle x_{\parallel max} \rangle+6\sigma$ (horizontal red dashed line) is defined to gauge the extent of the spike height and present a picture that they even cross this larger significant threshold height. Such occasional large events are often reported as extreme events in the literature \cite{mishra2020routes, chowdhury2022extreme}.  
\par The probability density distribution (PDF) of all the local peaks $x_{\parallel max}$ = $x_n$
shows an interesting distribution that follows a power law 
 against the height of spikes \cite{mishra2018dragon, mishra2020routes}. 
The low to moderate amplitude events follow a power law (a dotted straight line) while the larger events beyond the threshold (vertical dashed line) are outliers \cite{mishra2018dragon, mishra2020routes} as shown in a log-log plot of probability density against the height of events ($x_n$) in Fig.~\ref{HR:TIME_PDF}(b). The probability of large-amplitude bursting events increases with the formation of a dragon-king-like bump \cite{cavalcante2013predictability} beyond a threshold height of $h_{s_1}$ marked by a vertical line (dashed line). 
We  demonstrate  a discontinuous transition to hyperchaos from a periodic state in response to a parameter change when two Lyapunov exponents shift simultaneously to a positive value at a critical parameter. The  typical pattern of temporal dynamics appears with recurrent large-amplitude spikes and bursts with a dragon-king like distribution of heights of the spikes. 
We elaborate this scenario further in two more example systems in Secs. III and IV. 

\section{Three coupled duffing oscillator}
\begin{figure} 
	\begin{center}
		\includegraphics[width=0.85\columnwidth]{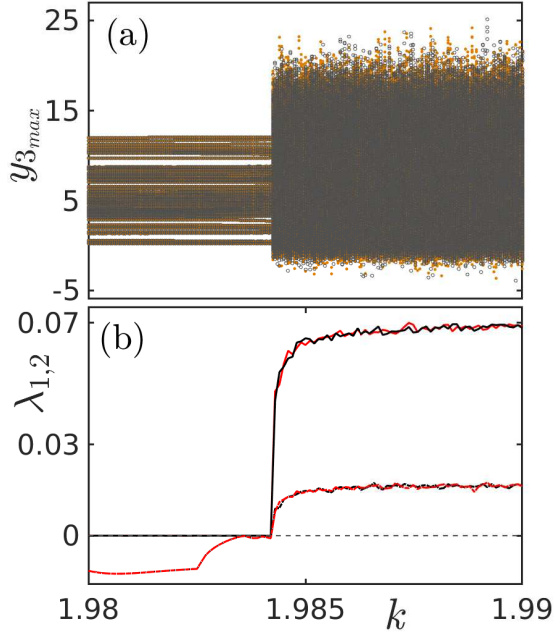}
	\end{center}
	\caption{Hyperchaos in a ring of three-coupled Duffing oscillators. (a) Bifurcation diagram of $y_3{_{max}}$ against the coupling strength $k$ that shows a sudden surge of  amplitude at critical control parameter $k$ = 1.9843, and (b) the two largest Lyapunov exponent shows a direct transition from a periodic state to hyperchaos when both $\lambda_{1,2}$ become  positive simultaneously. The emergence of hyperchaos and a rare large-amplitude oscillation takes place at a critical parameter value $k$ = 1.9843. Both forward and backward integrations in (b) (red and black lines) reveal a hysteresis-free transition to hyperchaos.}
	\label{DUF_BILYA} 
\end{figure}

Next we consider a model of three unidirectionally coupled Duffing oscillators \cite{perlikowski2010routes}, 
\begin{eqnarray}
	\dot{x_{1}}&=&y_{1}, \\
	\dot{y_{1}}&=&-dy_{1}-ax_{1}-x^3_{1}+k(x_{3}-x_{1}), \nonumber \\
	\dot{x_{2}}&=&y_{2}, \nonumber \\
	\dot{y_{2}}&=&-dy_{2}-ax_{2}-x^3_{2}+k(x_{1}-x_{2}), \nonumber \\
	\dot{x_{3}}&=&y_{3}, \nonumber \\
	\dot{y_{3}}&=&-dy_{3}-ax_{3}-x^3_{3}+k(x_{2}-x_{3}), \nonumber
\end{eqnarray}
where $k$ is the coupling strength. The system parameters are $a$ = 0.1, $d$ = 0.3, and the coupling is varied in a narrow range of $k \in (1.98, 1.99)$ when the ring of oscillators exhibits typical Pomeau-Manneville intermittency with almost periodic oscillations intercepted by occasional large-amplitude spikes. The  dynamics changes from a higher periodic oscillation to hyperchaos at a critical $k$ as illustrated in a bifurcation diagram in Fig.~\ref{DUF_BILYA}(a). 
A sudden large expansion in  $y_{3_{max}}$ appears  at a critical coupling strength $k$ = 1.9843 and it continues for larger values of coupling strength. The origin of hyperchaos from a periodic state is proved by a plot of two Lyapunov exponents $\lambda_{1,2}$  in Fig.~\ref{DUF_BILYA}(b). The largest Lyapunov exponent $\lambda_1$ remains zero and the second largest Lyapunov exponent $\lambda_2$ is negative for $k$ below a critical value $k = 1.9843$ when the dynamics is  periodic. Both the Lyapunov exponents simultaneously cross the zero line at critical $k$ = 1.9843, which exactly coincides with the sudden large expansion of $y_{3_{max}}$ as shown in the bifurcation diagram. A forward as well as backward integration of the model of the ring of oscillators is done which is reflected in two different colors (red and black lines) in Fig.~\ref{DUF_BILYA}(b), and it clearly exemplifies no shift in the transition point confirming  a hysteresis-free discontinuous transition to hyperchaos. 
\par
\begin{figure}
	\includegraphics[width=0.8\columnwidth]{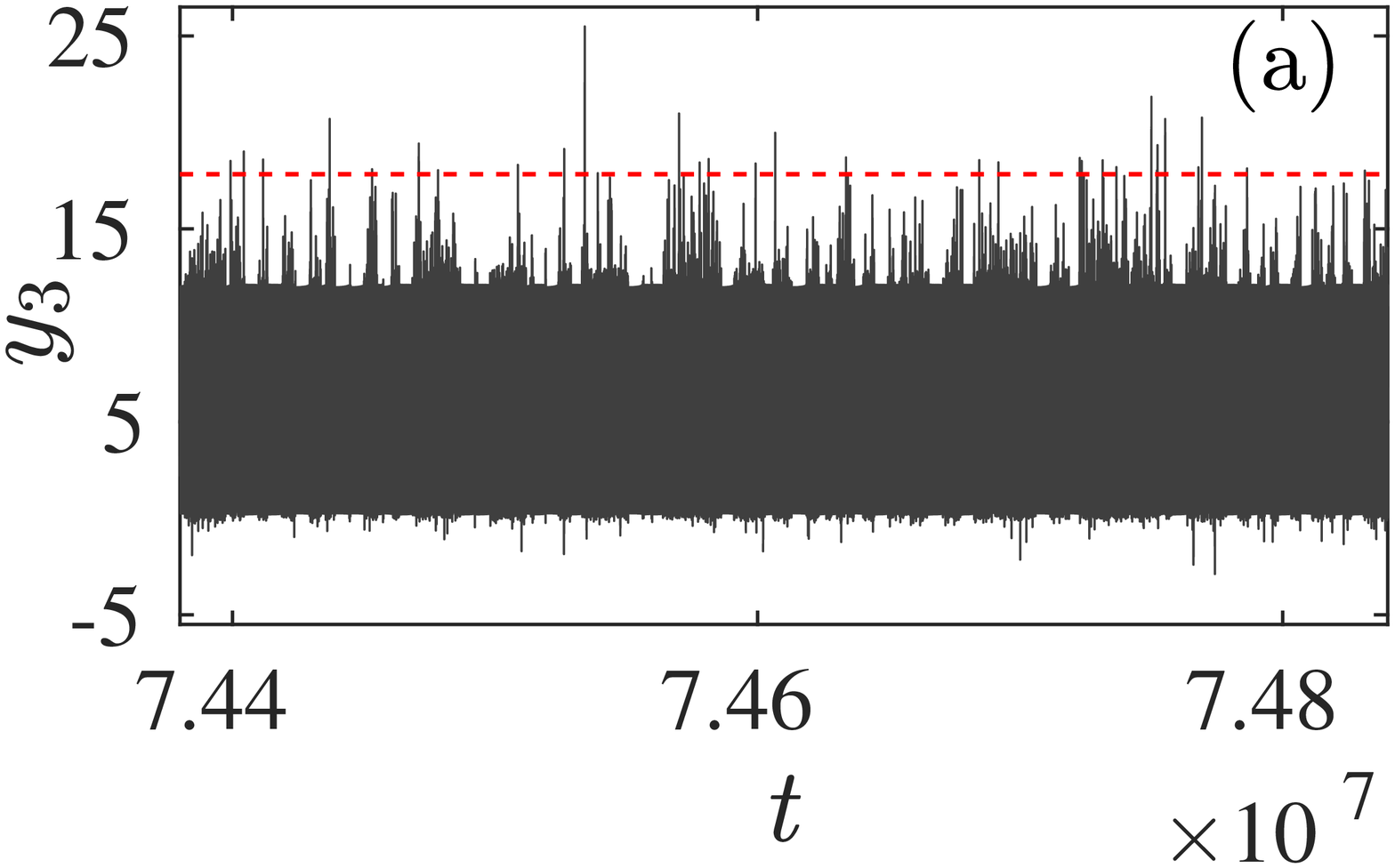}\\
	\includegraphics[height=4cm, width=4.55cm]{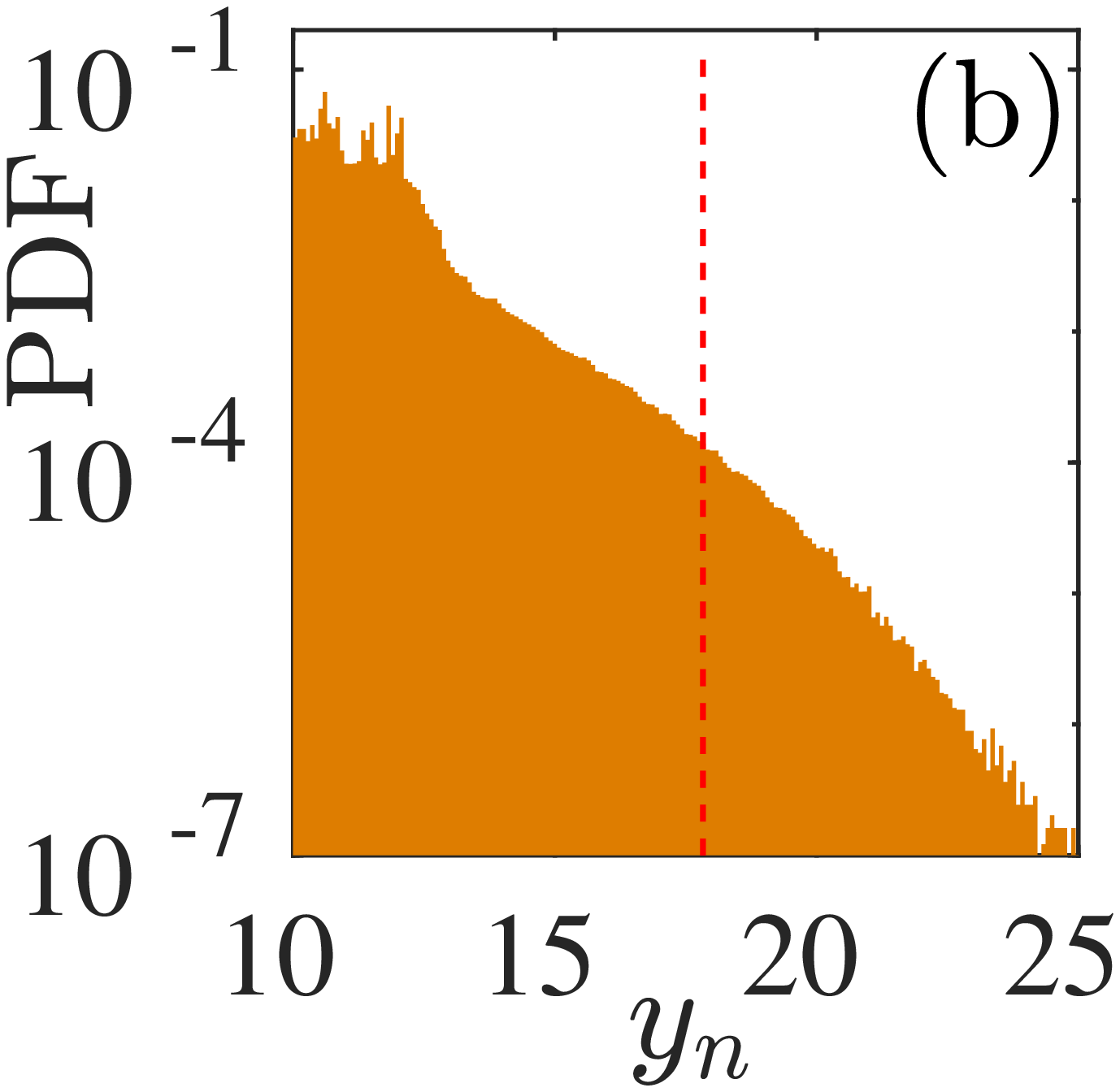}\\
	
	\caption{(a) Time evolution of hyperchaos in a ring of three-coupled Duffing oscillators at critical coupling strength $k$ = 1.9843. (b) PDF of $y_n$ illustrates a non-Gaussian distribution with a heavy tail  confirming rare occurrence of large events in (a) that exceed the significant height $h_s$ (horizontal dashed red line). }
	\label{DUF:TIME_PDF}
\end{figure}

 The time evolution of hyperchaos beyond a critical point is plotted in Fig.~\ref{DUF:TIME_PDF}(a), elucidating appearance of large-amplitude spikes. Some of the spikes are larger than  a threshold height $h_s=  \langle y_n\rangle$ + $4 \sigma$ ( horizontal dashed line), where $y_n=y_{3_{max}}$. The probability of occurrence of large spikes of amplitude $y_n$ of a different size as seen in the temporal dynamics shows a non-Gaussian heavy-tail distribution in Fig.~\ref{DUF:TIME_PDF}(b) where the amplitude of some spikes is really rare and large enough to cross the threshold height $h_s$ (vertical dashed line).  It is noteworthy to mention here that the earlier study \cite{perlikowski2010routes} using the same model investigated a different parameter region and did not focus on this narrow parameter region of $k \in (1.98, 1.99)$ and, therefore, did not observe the Pomeau-Manneville intermittency when the simultaneous shifting of two Lyapunov exponents occur at a critical value with the emergence of hyperchaos. They did not investigate the statistical feature of the temporal dynamics. The occasional large-amplitude events in the temporal dynamics clearly manifest a typical feature of extreme events. In the next example, we demonstrate that the transition to hyperchaos is not restricted to Pomeau-Manneville intermittency only.

\begin{figure}
	\begin{center}

		\includegraphics[width=0.82\columnwidth]{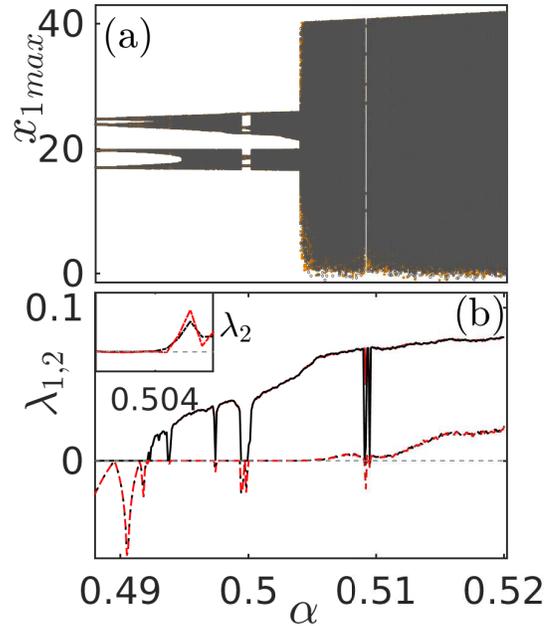}
	\end{center}
	\caption{Crisis-induced intermittency and appearance of hyperchaos in a four-dimensional circuit model. (a) Bifurcation diagram of $x_{1_{max}}$ and (b) its corroborate largest Lyapunov exponent against $\alpha$. Two different colors in (a) and (b) signify hysteresis-free transitions  near the critical point of $\alpha$ = 0.504. Inset shows an enlarged view of $\lambda_2$ plot at the critical point.}
	\label{CIR_BILYA} 
\end{figure}
\section{Hyperchaotic Oscillator with Gyrators}

\begin{figure}
	\includegraphics[width=0.82\columnwidth]{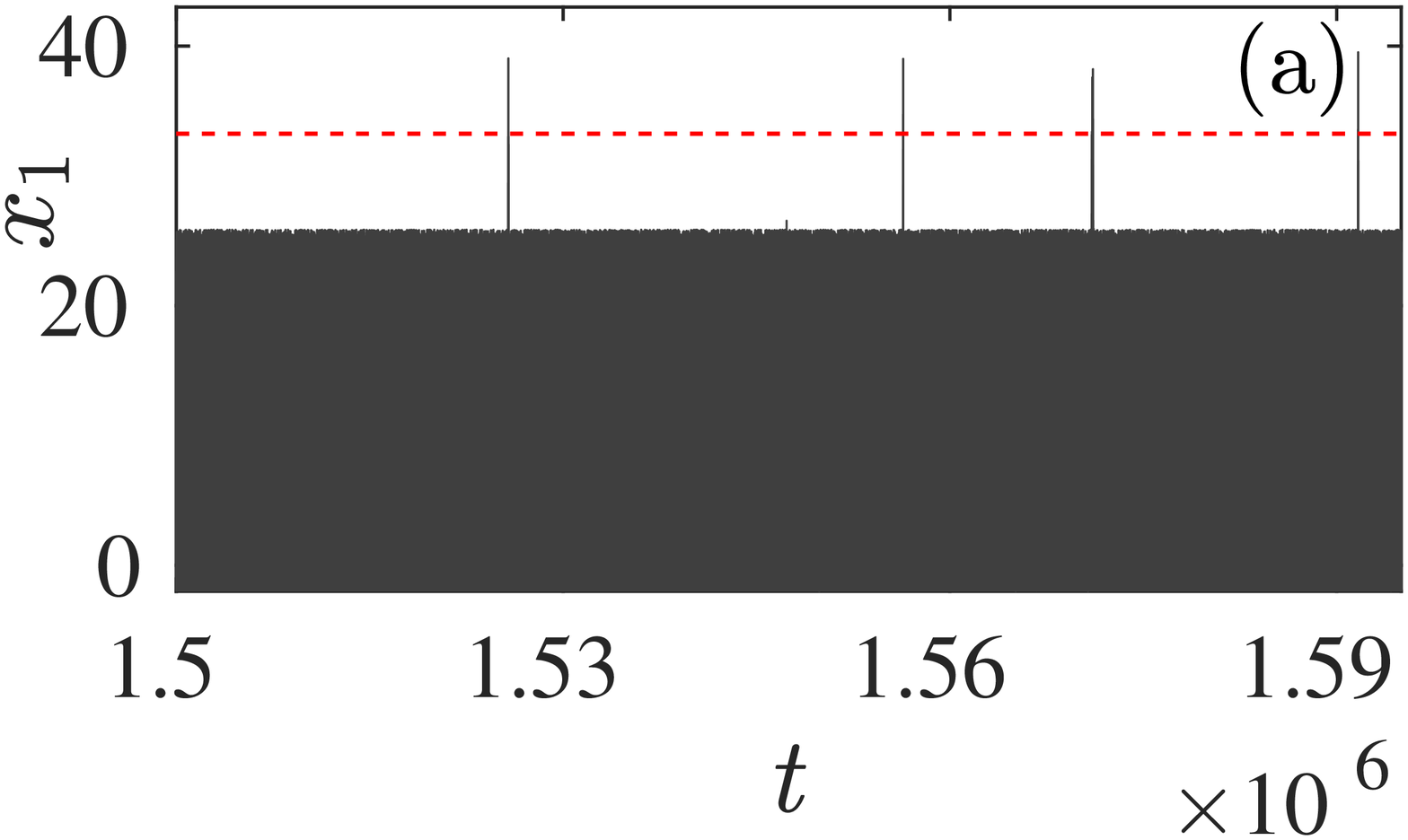}\\
	\includegraphics[width=0.52\columnwidth]{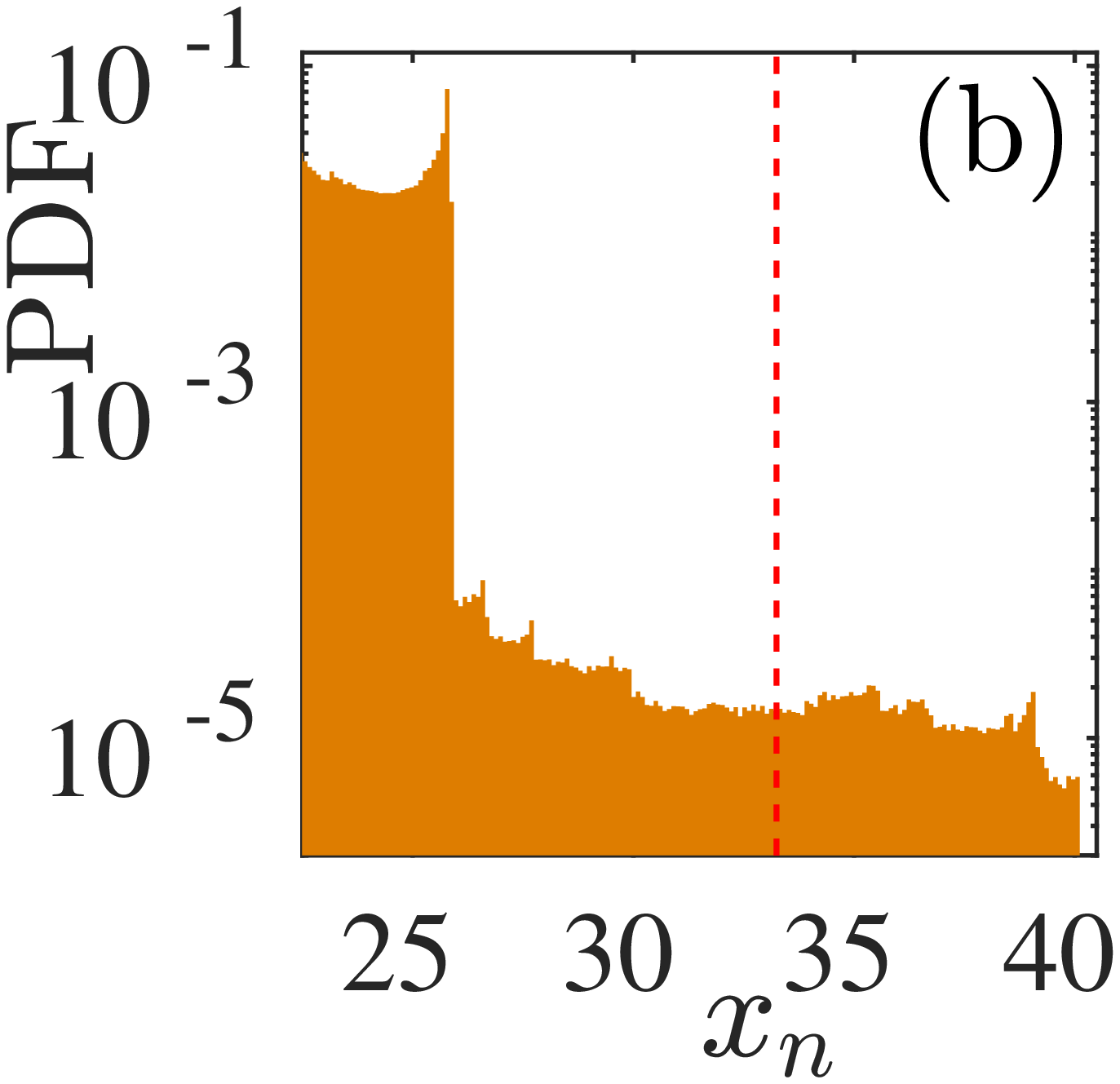}
	\caption{ Hyperchaotic circuit model. (a) Temporal dynamics of  hyperchaos for  $\alpha$ = 0.504. The dynamics remains confined to the dense region (black) but with rare large spikes larger than a threshold line $h_s$ (horizontal dashed line). (b) Probability of occurrence of $x_n=max(x_{1})$ shows a non-Gaussian distribution with a tail beyond the $h_s$ threshold (vertical dashed line).}	
	\label{CIR:TIME_PDF}
\end{figure}

Finally, we demonstrate the origin and temporal behavior of hyperchaos in a four-dimensional model \cite{kengne2015coexistence,tamasevicuius1997hyperchaotic}. The normalized form of the governing dynamics is given below:
\begin{eqnarray} 
\dot{x_1}&=&-\alpha x_1-x_2-\gamma \varphi(x_1 - x_3)  \\ 
\dot{x_2}&=&x_1 \nonumber \\ 
\dot{x_3}&=&-\varepsilon(-x_4 +\gamma \varphi(x_1 - x_3))  \nonumber \\ 
\dot{x_4}&=&\mu x_3 \nonumber 
\end{eqnarray}
where $\varphi(x_1 - x_3)=\mathrm{exp} ((x_1 - x_3)-1.0)$. The parameters $\gamma$ = $5.856531\times10^{-5}$,  $\varepsilon$ = 3.235294, and $\mu$ = 3.047619 are kept fixed while varying $\alpha \in (0.5, 0.515)$ and when the system first manifests a period-doubling route to chaos and then transits to hyperchaos via  crisis-induced intermittency \cite{grebogi1987critical}. 
The bifurcation of $x_1{_{max}}$ in Fig.~\ref{CIR_BILYA}(a) once again illustrates a discontinuous transition with a large expansion of the attractor at a critical $\alpha$ = 0.504. This indicates a transition to hyperchaos, as confirmed once again by the  transition of the second Lyapunov exponent $\lambda_2$ to a positive value at $\alpha$ = 0.504 as shown in Fig.~\ref{CIR_BILYA}(b).
~The transition point of hyperchaos at the critical parameter is not clear in Fig.~\ref{CIR_BILYA}(b) since $\lambda_1$ is very large; however, a zoomed version (see the insert) apparently manifests the transition point at the critical parameter when $\lambda_2$ moves to a positive value. Once again the results of forward and backward integrations of two different colors  are presented in Fig.~\ref{CIR_BILYA}(a) and \ref{CIR_BILYA}(b) that denote a hysteresis-free transition at a critical point. The temporal dynamics of hyperchaos shows a dark region of  nominal amplitude chaos, but with rare  large-amplitude spikes in Fig.~\ref{CIR:TIME_PDF}(a) which are larger than a threshold height  $h_s$= $ \langle x_n\rangle$ + $4 \sigma$ (red dashed line), where $x_n=x_{1_{max}}$. The rare spikes have crossed the threshold height $h_s$. PDF of all the peaks of $x_n$ is presented in Fig.~\ref{CIR:TIME_PDF}(b)  manifesting a non-Gaussian distribution with a tail and confirms a rare  occurrence of the large spikes, a typical feature of extreme events. 

\section{conclusion}

The origin of hyperchaos has been widely studied in many models and practical systems. It has been understood that hyperchaos may appear from a periodic, quasiperiodic, and chaotic state via different dynamical processes with the appearance of at least two positive Lyapunov exponents. We realize a common characteristic feature of hyperchaos from our study of three paradigmatic models, a coupled neuron model, a ring of three Duffing oscillators, and a hyperchaotic circuit model that confirms our earlier observation of the same in the Zeeman laser model. 
Looking first at the bifurcation diagram of the system dynamics, we noted that a discontinuous transition to hyperchaos appears at a critical parameter when a large expansion of the attractor of the system occurs. This is a common scenario for all the  varieties of original dynamics (periodic, quasiperiodic, and chaotic) from where the dynamics switches to hyperchaos in response to a parameter change. This transition may be hysteresis-free that depends upon the system. In two example cases, two Lyapunov exponents shift to a positive value simultaneously, and in another case, they shift to positive values one after another with a change in the parameter. 
The large expansion of the attractor has its manifestation in the temporal dynamics that is bounded in amplitude, most of the time, but shows a highly irregular and occasional occurrence of large-amplitude spikes that are larger than a threshold height.
The probability distribution of all the spikes (local maxima) in the temporal dynamics, in particular, shows a non-Gaussian statistics with a tail (heavy or light) that indicates a rare and recurrent occurrence of large-amplitude spikes. 
\par To summarize, we present a seemingly common picture of hyperchaos in the temporal dynamics of a system and in the long time statistics of local maxima. This observation is connected to a discontinuous transition to a large expansion of the attractor of a system, in general, during the origin of hyperchaos in response to a system parameter. The temporal pattern of hyperchaos shows recurrent and rare large events. Such a dynamical as well as a statistical characterization of hyperchaos is not reported earlier to the best of our knowledge. This unique behavior of hyperchaos is surely system dependent and also depends on the choice of parameters. In conclusion, we mention that hyperchaos is known as more complex than chaos; however, we raise a fundamental question here as to how we can classify hyperchaos with intermittent large-amplitude events, which are visibly more complex than hyperchaos without the presence of them. It is our future desire to understand this complexity with a possible quantitative measure.

\begin{acknowledgments}
	T.K. and S.L.K. have been supported by the National Science Centre, Poland, OPUS Programs (Project Nos.~2018/29/B/ST8/00457 and 2021/43/B/ST8/00641). S.K.D. is supported by the Division of Dynamics, Lodz University of Technology, Poland.
\end{acknowledgments}     
\section*{Data Availability}
The data that support the findings of this study are available from the corresponding author upon reasonable request.
\section*{References}
\bibliography{hyper_ref}

\begin{thebibliography}{31}%
\makeatletter
\providecommand \@ifxundefined [1]{%
 \@ifx{#1\undefined}
}%
\providecommand \@ifnum [1]{%
 \ifnum #1\expandafter \@firstoftwo
 \else \expandafter \@secondoftwo
 \fi
}%
\providecommand \@ifx [1]{%
 \ifx #1\expandafter \@firstoftwo
 \else \expandafter \@secondoftwo
 \fi
}%
\providecommand \natexlab [1]{#1}%
\providecommand \enquote  [1]{``#1''}%
\providecommand \bibnamefont  [1]{#1}%
\providecommand \bibfnamefont [1]{#1}%
\providecommand \citenamefont [1]{#1}%
\providecommand \href@noop [0]{\@secondoftwo}%
\providecommand \href [0]{\begingroup \@sanitize@url \@href}%
\providecommand \@href[1]{\@@startlink{#1}\@@href}%
\providecommand \@@href[1]{\endgroup#1\@@endlink}%
\providecommand \@sanitize@url [0]{\catcode `\\12\catcode `\$12\catcode
  `\&12\catcode `\#12\catcode `\^12\catcode `\_12\catcode `\%12\relax}%
\providecommand \@@startlink[1]{}%
\providecommand \@@endlink[0]{}%
\providecommand \url  [0]{\begingroup\@sanitize@url \@url }%
\providecommand \@url [1]{\endgroup\@href {#1}{\urlprefix }}%
\providecommand \urlprefix  [0]{URL }%
\providecommand \Eprint [0]{\href }%
\providecommand \doibase [0]{http://dx.doi.org/}%
\providecommand \selectlanguage [0]{\@gobble}%
\providecommand \bibinfo  [0]{\@secondoftwo}%
\providecommand \bibfield  [0]{\@secondoftwo}%
\providecommand \translation [1]{[#1]}%
\providecommand \BibitemOpen [0]{}%
\providecommand \bibitemStop [0]{}%
\providecommand \bibitemNoStop [0]{.\EOS\space}%
\providecommand \EOS [0]{\spacefactor3000\relax}%
\providecommand \BibitemShut  [1]{\csname bibitem#1\endcsname}%
\let\auto@bib@innerbib\@empty
\bibitem [{\citenamefont {Poincar{\'e}}(1967)}]{poincare1967new}%
  \BibitemOpen
  \bibfield  {author} {\bibinfo {author} {\bibfnamefont {H.}~\bibnamefont
  {Poincar{\'e}}},\ }\href@noop {} {\emph {\bibinfo {title} {New methods of
  celestial mechanics}}},\ Vol.~\bibinfo {volume} {2}\ (\bibinfo  {publisher}
  {National Aeronautics and Space Administration},\ \bibinfo {year}
  {1967})\BibitemShut {NoStop}%
\bibitem [{\citenamefont {Lorenz}(1963)}]{lorenz1963deterministic}%
  \BibitemOpen
  \bibfield  {author} {\bibinfo {author} {\bibfnamefont {E.~N.}\ \bibnamefont
  {Lorenz}},\ }\bibfield  {title} {\enquote {\bibinfo {title} {Deterministic
  nonperiodic flow},}\ }\href@noop {} {\bibfield  {journal} {\bibinfo
  {journal} {Journal of atmospheric sciences}\ }\textbf {\bibinfo {volume}
  {20}},\ \bibinfo {pages} {130--141} (\bibinfo {year} {1963})}\BibitemShut
  {NoStop}%
\bibitem [{\citenamefont {Li}\ and\ \citenamefont
  {Yorke}(2004)}]{li2004period}%
  \BibitemOpen
  \bibfield  {author} {\bibinfo {author} {\bibfnamefont {T.-Y.}\ \bibnamefont
  {Li}}\ and\ \bibinfo {author} {\bibfnamefont {J.~A.}\ \bibnamefont {Yorke}},\
  }\bibfield  {title} {\enquote {\bibinfo {title} {Period three implies
  chaos},}\ }in\ \href@noop {} {\emph {\bibinfo {booktitle} {The theory of
  chaotic attractors}}}\ (\bibinfo  {publisher} {Springer},\ \bibinfo {year}
  {2004})\ pp.\ \bibinfo {pages} {77--84}\BibitemShut {NoStop}%
\bibitem [{\citenamefont {Alligood}, \citenamefont {Sauer},\ and\ \citenamefont
  {Yorke}(1996)}]{alligood1996chaos}%
  \BibitemOpen
  \bibfield  {author} {\bibinfo {author} {\bibfnamefont {K.~T.}\ \bibnamefont
  {Alligood}}, \bibinfo {author} {\bibfnamefont {T.~D.}\ \bibnamefont {Sauer}},
  \ and\ \bibinfo {author} {\bibfnamefont {J.~A.}\ \bibnamefont {Yorke}},\
  }\href@noop {} {\emph {\bibinfo {title} {Chaos}}}\ (\bibinfo  {publisher}
  {Springer},\ \bibinfo {year} {1996})\BibitemShut {NoStop}%
\bibitem [{\citenamefont {Rossler}(1979)}]{rossler1979equation}%
  \BibitemOpen
  \bibfield  {author} {\bibinfo {author} {\bibfnamefont {O.}~\bibnamefont
  {Rossler}},\ }\bibfield  {title} {\enquote {\bibinfo {title} {An equation for
  hyperchaos},}\ }\href@noop {} {\bibfield  {journal} {\bibinfo  {journal}
  {Physics Letters A}\ }\textbf {\bibinfo {volume} {71}},\ \bibinfo {pages}
  {155--157} (\bibinfo {year} {1979})}\BibitemShut {NoStop}%
\bibitem [{\citenamefont {Franco}, \citenamefont {Rempel},\ and\ \citenamefont
  {Munoz}(2020)}]{franco2020crisis}%
  \BibitemOpen
  \bibfield  {author} {\bibinfo {author} {\bibfnamefont {F.~F.}\ \bibnamefont
  {Franco}}, \bibinfo {author} {\bibfnamefont {E.~L.}\ \bibnamefont {Rempel}},
  \ and\ \bibinfo {author} {\bibfnamefont {P.~R.}\ \bibnamefont {Munoz}},\
  }\bibfield  {title} {\enquote {\bibinfo {title} {Crisis and hyperchaos in a
  simplified model of magnetoconvection},}\ }\href@noop {} {\bibfield
  {journal} {\bibinfo  {journal} {Physica D: Nonlinear Phenomena}\ }\textbf
  {\bibinfo {volume} {406}},\ \bibinfo {pages} {132417} (\bibinfo {year}
  {2020})}\BibitemShut {NoStop}%
\bibitem [{\citenamefont {Miranda}\ \emph {et~al.}(2013)\citenamefont
  {Miranda}, \citenamefont {Rempel}, \citenamefont {Chian}, \citenamefont
  {Seehafer}, \citenamefont {Toledo},\ and\ \citenamefont
  {Munoz}}]{miranda2013lagrangian}%
  \BibitemOpen
  \bibfield  {author} {\bibinfo {author} {\bibfnamefont {R.~A.}\ \bibnamefont
  {Miranda}}, \bibinfo {author} {\bibfnamefont {E.~L.}\ \bibnamefont {Rempel}},
  \bibinfo {author} {\bibfnamefont {A.~C.-L.}\ \bibnamefont {Chian}}, \bibinfo
  {author} {\bibfnamefont {N.}~\bibnamefont {Seehafer}}, \bibinfo {author}
  {\bibfnamefont {B.~A.}\ \bibnamefont {Toledo}}, \ and\ \bibinfo {author}
  {\bibfnamefont {P.~R.}\ \bibnamefont {Munoz}},\ }\bibfield  {title} {\enquote
  {\bibinfo {title} {Lagrangian coherent structures at the onset of hyperchaos
  in the two-dimensional navier-stokes equations},}\ }\href@noop {} {\bibfield
  {journal} {\bibinfo  {journal} {Chaos: An Interdisciplinary Journal of
  Nonlinear Science}\ }\textbf {\bibinfo {volume} {23}},\ \bibinfo {pages}
  {033107} (\bibinfo {year} {2013})}\BibitemShut {NoStop}%
\bibitem [{\citenamefont {Kapitaniak}(1993)}]{kapitaniak1993transition}%
  \BibitemOpen
  \bibfield  {author} {\bibinfo {author} {\bibfnamefont {T.}~\bibnamefont
  {Kapitaniak}},\ }\bibfield  {title} {\enquote {\bibinfo {title} {Transition
  to hyperchaos in chaotically forced coupled oscillators},}\ }\href@noop {}
  {\bibfield  {journal} {\bibinfo  {journal} {Physical Review E}\ }\textbf
  {\bibinfo {volume} {47}},\ \bibinfo {pages} {R2975} (\bibinfo {year}
  {1993})}\BibitemShut {NoStop}%
\bibitem [{\citenamefont {Perlikowski}\ \emph {et~al.}(2010)\citenamefont
  {Perlikowski}, \citenamefont {Yanchuk}, \citenamefont {Wolfrum},
  \citenamefont {Stefanski}, \citenamefont {Mosiolek},\ and\ \citenamefont
  {Kapitaniak}}]{perlikowski2010routes}%
  \BibitemOpen
  \bibfield  {author} {\bibinfo {author} {\bibfnamefont {P.}~\bibnamefont
  {Perlikowski}}, \bibinfo {author} {\bibfnamefont {S.}~\bibnamefont
  {Yanchuk}}, \bibinfo {author} {\bibfnamefont {M.}~\bibnamefont {Wolfrum}},
  \bibinfo {author} {\bibfnamefont {A.}~\bibnamefont {Stefanski}}, \bibinfo
  {author} {\bibfnamefont {P.}~\bibnamefont {Mosiolek}}, \ and\ \bibinfo
  {author} {\bibfnamefont {T.}~\bibnamefont {Kapitaniak}},\ }\bibfield  {title}
  {\enquote {\bibinfo {title} {Routes to complex dynamics in a ring of
  unidirectionally coupled systems},}\ }\href@noop {} {\bibfield  {journal}
  {\bibinfo  {journal} {Chaos: An Interdisciplinary Journal of Nonlinear
  Science}\ }\textbf {\bibinfo {volume} {20}},\ \bibinfo {pages} {013111}
  (\bibinfo {year} {2010})}\BibitemShut {NoStop}%
\bibitem [{\citenamefont {Kengne}(2015)}]{kengne2015coexistence}%
  \BibitemOpen
  \bibfield  {author} {\bibinfo {author} {\bibfnamefont {J.}~\bibnamefont
  {Kengne}},\ }\bibfield  {title} {\enquote {\bibinfo {title} {Coexistence of
  chaos with hyperchaos, period-3 doubling bifurcation, and transient chaos in
  the hyperchaotic oscillator with gyrators},}\ }\href@noop {} {\bibfield
  {journal} {\bibinfo  {journal} {International Journal of Bifurcation and
  Chaos}\ }\textbf {\bibinfo {volume} {25}},\ \bibinfo {pages} {1550052}
  (\bibinfo {year} {2015})}\BibitemShut {NoStop}%
\bibitem [{\citenamefont {Tamasevicuius}\ \emph {et~al.}(1997)\citenamefont
  {Tamasevicuius}, \citenamefont {Cenys}, \citenamefont {Mykolaitis},
  \citenamefont {Namajunas},\ and\ \citenamefont
  {Lindberg}}]{tamasevicuius1997hyperchaotic}%
  \BibitemOpen
  \bibfield  {author} {\bibinfo {author} {\bibfnamefont {A.}~\bibnamefont
  {Tamasevicuius}}, \bibinfo {author} {\bibfnamefont {A.}~\bibnamefont
  {Cenys}}, \bibinfo {author} {\bibfnamefont {G.}~\bibnamefont {Mykolaitis}},
  \bibinfo {author} {\bibfnamefont {A.}~\bibnamefont {Namajunas}}, \ and\
  \bibinfo {author} {\bibfnamefont {E.}~\bibnamefont {Lindberg}},\ }\bibfield
  {title} {\enquote {\bibinfo {title} {Hyperchaotic oscillator with
  gyrators},}\ }\href@noop {} {\bibfield  {journal} {\bibinfo  {journal}
  {Electronics Letters}\ }\textbf {\bibinfo {volume} {33}},\ \bibinfo {pages}
  {542--544} (\bibinfo {year} {1997})}\BibitemShut {NoStop}%
\bibitem [{\citenamefont {Colet}, \citenamefont {Roy},\ and\ \citenamefont
  {Wiesenfeld}(1994)}]{colet1994controlling}%
  \BibitemOpen
  \bibfield  {author} {\bibinfo {author} {\bibfnamefont {P.}~\bibnamefont
  {Colet}}, \bibinfo {author} {\bibfnamefont {R.}~\bibnamefont {Roy}}, \ and\
  \bibinfo {author} {\bibfnamefont {K.}~\bibnamefont {Wiesenfeld}},\ }\bibfield
   {title} {\enquote {\bibinfo {title} {Controlling hyperchaos in a multimode
  laser model},}\ }\href@noop {} {\bibfield  {journal} {\bibinfo  {journal}
  {Physical Review E}\ }\textbf {\bibinfo {volume} {50}},\ \bibinfo {pages}
  {3453} (\bibinfo {year} {1994})}\BibitemShut {NoStop}%
\bibitem [{\citenamefont {Stankevich}\ \emph {et~al.}(2018)\citenamefont
  {Stankevich}, \citenamefont {Dvorak}, \citenamefont {Astakhov}, \citenamefont
  {Jaros}, \citenamefont {Kapitaniak}, \citenamefont {Perlikowski},\ and\
  \citenamefont {Kapitaniak}}]{stankevich2018chaos}%
  \BibitemOpen
  \bibfield  {author} {\bibinfo {author} {\bibfnamefont {N.~V.}\ \bibnamefont
  {Stankevich}}, \bibinfo {author} {\bibfnamefont {A.}~\bibnamefont {Dvorak}},
  \bibinfo {author} {\bibfnamefont {V.}~\bibnamefont {Astakhov}}, \bibinfo
  {author} {\bibfnamefont {P.}~\bibnamefont {Jaros}}, \bibinfo {author}
  {\bibfnamefont {M.}~\bibnamefont {Kapitaniak}}, \bibinfo {author}
  {\bibfnamefont {P.}~\bibnamefont {Perlikowski}}, \ and\ \bibinfo {author}
  {\bibfnamefont {T.}~\bibnamefont {Kapitaniak}},\ }\bibfield  {title}
  {\enquote {\bibinfo {title} {Chaos and hyperchaos in coupled antiphase driven
  toda oscillators},}\ }\href@noop {} {\bibfield  {journal} {\bibinfo
  {journal} {Regular and Chaotic Dynamics}\ }\textbf {\bibinfo {volume} {23}},\
  \bibinfo {pages} {120--126} (\bibinfo {year} {2018})}\BibitemShut {NoStop}%
\bibitem [{\citenamefont {Stankevich}\ \emph {et~al.}(2019)\citenamefont
  {Stankevich}, \citenamefont {Kuznetsov}, \citenamefont {Popova},\ and\
  \citenamefont {Seleznev}}]{stankevich2019chaos}%
  \BibitemOpen
  \bibfield  {author} {\bibinfo {author} {\bibfnamefont {N.}~\bibnamefont
  {Stankevich}}, \bibinfo {author} {\bibfnamefont {A.}~\bibnamefont
  {Kuznetsov}}, \bibinfo {author} {\bibfnamefont {E.}~\bibnamefont {Popova}}, \
  and\ \bibinfo {author} {\bibfnamefont {E.}~\bibnamefont {Seleznev}},\
  }\bibfield  {title} {\enquote {\bibinfo {title} {Chaos and hyperchaos via
  secondary neimark--sacker bifurcation in a model of radiophysical
  generator},}\ }\href@noop {} {\bibfield  {journal} {\bibinfo  {journal}
  {Nonlinear dynamics}\ }\textbf {\bibinfo {volume} {97}},\ \bibinfo {pages}
  {2355--2370} (\bibinfo {year} {2019})}\BibitemShut {NoStop}%
\bibitem [{\citenamefont {Kashyap}\ and\ \citenamefont
  {Kolwankar}(2020)}]{kashyap2020hyperchaos}%
  \BibitemOpen
  \bibfield  {author} {\bibinfo {author} {\bibfnamefont {A.~R.}\ \bibnamefont
  {Kashyap}}\ and\ \bibinfo {author} {\bibfnamefont {K.~M.}\ \bibnamefont
  {Kolwankar}},\ }\bibfield  {title} {\enquote {\bibinfo {title} {Hyperchaos
  and synchronization in two element nonlinear chimney model},}\ }\href@noop {}
  {\bibfield  {journal} {\bibinfo  {journal} {Chaos: An Interdisciplinary
  Journal of Nonlinear Science}\ }\textbf {\bibinfo {volume} {30}},\ \bibinfo
  {pages} {123114} (\bibinfo {year} {2020})}\BibitemShut {NoStop}%
\bibitem [{\citenamefont {Stankevich}\ and\ \citenamefont
  {Volkov}(2021)}]{stankevich2021chaos}%
  \BibitemOpen
  \bibfield  {author} {\bibinfo {author} {\bibfnamefont {N.}~\bibnamefont
  {Stankevich}}\ and\ \bibinfo {author} {\bibfnamefont {E.}~\bibnamefont
  {Volkov}},\ }\bibfield  {title} {\enquote {\bibinfo {title}
  {Chaos--hyperchaos transition in three identical quorum-sensing mean-field
  coupled ring oscillators},}\ }\href@noop {} {\bibfield  {journal} {\bibinfo
  {journal} {Chaos: An Interdisciplinary Journal of Nonlinear Science}\
  }\textbf {\bibinfo {volume} {31}},\ \bibinfo {pages} {103112} (\bibinfo
  {year} {2021})}\BibitemShut {NoStop}%
\bibitem [{\citenamefont {Garashchuk}\ \emph {et~al.}(2019)\citenamefont
  {Garashchuk}, \citenamefont {Sinelshchikov}, \citenamefont {Kazakov},\ and\
  \citenamefont {Kudryashov}}]{garashchuk2019hyperchaos}%
  \BibitemOpen
  \bibfield  {author} {\bibinfo {author} {\bibfnamefont {I.~R.}\ \bibnamefont
  {Garashchuk}}, \bibinfo {author} {\bibfnamefont {D.~I.}\ \bibnamefont
  {Sinelshchikov}}, \bibinfo {author} {\bibfnamefont {A.~O.}\ \bibnamefont
  {Kazakov}}, \ and\ \bibinfo {author} {\bibfnamefont {N.~A.}\ \bibnamefont
  {Kudryashov}},\ }\bibfield  {title} {\enquote {\bibinfo {title} {Hyperchaos
  and multistability in the model of two interacting microbubble contrast
  agents},}\ }\href@noop {} {\bibfield  {journal} {\bibinfo  {journal} {Chaos:
  An Interdisciplinary Journal of Nonlinear Science}\ }\textbf {\bibinfo
  {volume} {29}},\ \bibinfo {pages} {063131} (\bibinfo {year}
  {2019})}\BibitemShut {NoStop}%
\bibitem [{\citenamefont {Stankevich}, \citenamefont {Kazakov},\ and\
  \citenamefont {Gonchenko}(2020)}]{stankevich2020scenarios}%
  \BibitemOpen
  \bibfield  {author} {\bibinfo {author} {\bibfnamefont {N.}~\bibnamefont
  {Stankevich}}, \bibinfo {author} {\bibfnamefont {A.}~\bibnamefont {Kazakov}},
  \ and\ \bibinfo {author} {\bibfnamefont {S.}~\bibnamefont {Gonchenko}},\
  }\bibfield  {title} {\enquote {\bibinfo {title} {Scenarios of hyperchaos
  occurrence in 4d r{\"o}ssler system},}\ }\href@noop {} {\bibfield  {journal}
  {\bibinfo  {journal} {Chaos: An Interdisciplinary Journal of Nonlinear
  Science}\ }\textbf {\bibinfo {volume} {30}},\ \bibinfo {pages} {123129}
  (\bibinfo {year} {2020})}\BibitemShut {NoStop}%
\bibitem [{\citenamefont {Bonatto}(2018)}]{bonatto2018hyperchaotic}%
  \BibitemOpen
  \bibfield  {author} {\bibinfo {author} {\bibfnamefont {C.}~\bibnamefont
  {Bonatto}},\ }\bibfield  {title} {\enquote {\bibinfo {title} {Hyperchaotic
  dynamics for light polarization in a laser diode},}\ }\href@noop {}
  {\bibfield  {journal} {\bibinfo  {journal} {Physical review letters}\
  }\textbf {\bibinfo {volume} {120}},\ \bibinfo {pages} {163902} (\bibinfo
  {year} {2018})}\BibitemShut {NoStop}%
\bibitem [{\citenamefont {Momp{\'o}}, \citenamefont {Carretero},\ and\
  \citenamefont {Bonilla}(2021)}]{mompo2021designing}%
  \BibitemOpen
  \bibfield  {author} {\bibinfo {author} {\bibfnamefont {E.}~\bibnamefont
  {Momp{\'o}}}, \bibinfo {author} {\bibfnamefont {M.}~\bibnamefont
  {Carretero}}, \ and\ \bibinfo {author} {\bibfnamefont {L.}~\bibnamefont
  {Bonilla}},\ }\bibfield  {title} {\enquote {\bibinfo {title} {Designing
  hyperchaos and intermittency in semiconductor superlattices},}\ }\href@noop
  {} {\bibfield  {journal} {\bibinfo  {journal} {Physical Review Letters}\
  }\textbf {\bibinfo {volume} {127}},\ \bibinfo {pages} {096601} (\bibinfo
  {year} {2021})}\BibitemShut {NoStop}%
\bibitem [{\citenamefont {Kingston}\ \emph {et~al.}(2022)\citenamefont
  {Kingston}, \citenamefont {Balcerzak}, \citenamefont {Kapitaniak},\ and\
  \citenamefont {Dana}}]{kingston2022transition}%
  \BibitemOpen
  \bibfield  {author} {\bibinfo {author} {\bibfnamefont {S.~L.}\ \bibnamefont
  {Kingston}}, \bibinfo {author} {\bibfnamefont {M.}~\bibnamefont {Balcerzak}},
  \bibinfo {author} {\bibfnamefont {T.}~\bibnamefont {Kapitaniak}}, \ and\
  \bibinfo {author} {\bibfnamefont {S.~K.}\ \bibnamefont {Dana}},\ }\bibfield
  {title} {\enquote {\bibinfo {title} {Transition to hyperchaos and rare
  large-intensity pulses in zeeman laser},}\ }\href@noop {} {\bibfield
  {journal} {\bibinfo  {journal} {arXiv preprint arXiv:2201.09567}\ } (\bibinfo
  {year} {2022})}\BibitemShut {NoStop}%
\bibitem [{\citenamefont {Pomeau}\ and\ \citenamefont
  {Manneville}(1980)}]{pomeau1980intermittent}%
  \BibitemOpen
  \bibfield  {author} {\bibinfo {author} {\bibfnamefont {Y.}~\bibnamefont
  {Pomeau}}\ and\ \bibinfo {author} {\bibfnamefont {P.}~\bibnamefont
  {Manneville}},\ }\bibfield  {title} {\enquote {\bibinfo {title} {Intermittent
  transition to turbulence in dissipative dynamical systems},}\ }\href@noop {}
  {\bibfield  {journal} {\bibinfo  {journal} {Communications in Mathematical
  Physics}\ }\textbf {\bibinfo {volume} {74}},\ \bibinfo {pages} {189--197}
  (\bibinfo {year} {1980})}\BibitemShut {NoStop}%
\bibitem [{\citenamefont {Kingston}\ \emph {et~al.}(2017)\citenamefont
  {Kingston}, \citenamefont {Thamilmaran}, \citenamefont {Pal}, \citenamefont
  {Feudel},\ and\ \citenamefont {Dana}}]{kingston2017extreme}%
  \BibitemOpen
  \bibfield  {author} {\bibinfo {author} {\bibfnamefont {S.~L.}\ \bibnamefont
  {Kingston}}, \bibinfo {author} {\bibfnamefont {K.}~\bibnamefont
  {Thamilmaran}}, \bibinfo {author} {\bibfnamefont {P.}~\bibnamefont {Pal}},
  \bibinfo {author} {\bibfnamefont {U.}~\bibnamefont {Feudel}}, \ and\ \bibinfo
  {author} {\bibfnamefont {S.~K.}\ \bibnamefont {Dana}},\ }\bibfield  {title}
  {\enquote {\bibinfo {title} {Extreme events in the forced li{\'e}nard
  system},}\ }\href@noop {} {\bibfield  {journal} {\bibinfo  {journal}
  {Physical Review E}\ }\textbf {\bibinfo {volume} {96}},\ \bibinfo {pages}
  {052204} (\bibinfo {year} {2017})}\BibitemShut {NoStop}%
\bibitem [{\citenamefont {Redondo}, \citenamefont {de~Valc{\'a}rcel},\ and\
  \citenamefont {Rold{\'a}n}(1997)}]{redondo1997intermittent}%
  \BibitemOpen
  \bibfield  {author} {\bibinfo {author} {\bibfnamefont {J.}~\bibnamefont
  {Redondo}}, \bibinfo {author} {\bibfnamefont {G.~J.}\ \bibnamefont
  {de~Valc{\'a}rcel}}, \ and\ \bibinfo {author} {\bibfnamefont
  {E.}~\bibnamefont {Rold{\'a}n}},\ }\bibfield  {title} {\enquote {\bibinfo
  {title} {Intermittent and quasiperiodic behavior in a zeeman laser model with
  large cavity anisotropy},}\ }\href@noop {} {\bibfield  {journal} {\bibinfo
  {journal} {Physical Review E}\ }\textbf {\bibinfo {volume} {56}},\ \bibinfo
  {pages} {6589} (\bibinfo {year} {1997})}\BibitemShut {NoStop}%
\bibitem [{\citenamefont {Kingston}\ \emph {et~al.}(2021)\citenamefont
  {Kingston}, \citenamefont {Mishra}, \citenamefont {Balcerzak}, \citenamefont
  {Kapitaniak},\ and\ \citenamefont {Dana}}]{kingston2021instabilities}%
  \BibitemOpen
  \bibfield  {author} {\bibinfo {author} {\bibfnamefont {S.~L.}\ \bibnamefont
  {Kingston}}, \bibinfo {author} {\bibfnamefont {A.}~\bibnamefont {Mishra}},
  \bibinfo {author} {\bibfnamefont {M.}~\bibnamefont {Balcerzak}}, \bibinfo
  {author} {\bibfnamefont {T.}~\bibnamefont {Kapitaniak}}, \ and\ \bibinfo
  {author} {\bibfnamefont {S.~K.}\ \bibnamefont {Dana}},\ }\bibfield  {title}
  {\enquote {\bibinfo {title} {Instabilities in quasiperiodic motion lead to
  intermittent large-intensity events in zeeman laser},}\ }\href@noop {}
  {\bibfield  {journal} {\bibinfo  {journal} {Physical Review E}\ }\textbf
  {\bibinfo {volume} {104}},\ \bibinfo {pages} {034215} (\bibinfo {year}
  {2021})}\BibitemShut {NoStop}%
\bibitem [{\citenamefont {Grebogi}\ \emph {et~al.}(1987)\citenamefont
  {Grebogi}, \citenamefont {Ott}, \citenamefont {Romeiras},\ and\ \citenamefont
  {Yorke}}]{grebogi1987critical}%
  \BibitemOpen
  \bibfield  {author} {\bibinfo {author} {\bibfnamefont {C.}~\bibnamefont
  {Grebogi}}, \bibinfo {author} {\bibfnamefont {E.}~\bibnamefont {Ott}},
  \bibinfo {author} {\bibfnamefont {F.}~\bibnamefont {Romeiras}}, \ and\
  \bibinfo {author} {\bibfnamefont {J.~A.}\ \bibnamefont {Yorke}},\ }\bibfield
  {title} {\enquote {\bibinfo {title} {Critical exponents for crisis-induced
  intermittency},}\ }\href@noop {} {\bibfield  {journal} {\bibinfo  {journal}
  {Physical Review A}\ }\textbf {\bibinfo {volume} {36}},\ \bibinfo {pages}
  {5365} (\bibinfo {year} {1987})}\BibitemShut {NoStop}%
\bibitem [{\citenamefont {Mishra}\ \emph {et~al.}(2020)\citenamefont {Mishra},
  \citenamefont {Leo~Kingston}, \citenamefont {Hens}, \citenamefont
  {Kapitaniak}, \citenamefont {Feudel},\ and\ \citenamefont
  {Dana}}]{mishra2020routes}%
  \BibitemOpen
  \bibfield  {author} {\bibinfo {author} {\bibfnamefont {A.}~\bibnamefont
  {Mishra}}, \bibinfo {author} {\bibfnamefont {S.}~\bibnamefont
  {Leo~Kingston}}, \bibinfo {author} {\bibfnamefont {C.}~\bibnamefont {Hens}},
  \bibinfo {author} {\bibfnamefont {T.}~\bibnamefont {Kapitaniak}}, \bibinfo
  {author} {\bibfnamefont {U.}~\bibnamefont {Feudel}}, \ and\ \bibinfo {author}
  {\bibfnamefont {S.~K.}\ \bibnamefont {Dana}},\ }\bibfield  {title} {\enquote
  {\bibinfo {title} {Routes to extreme events in dynamical systems: Dynamical
  and statistical characteristics},}\ }\href@noop {} {\bibfield  {journal}
  {\bibinfo  {journal} {Chaos: An Interdisciplinary Journal of Nonlinear
  Science}\ }\textbf {\bibinfo {volume} {30}},\ \bibinfo {pages} {063114}
  (\bibinfo {year} {2020})}\BibitemShut {NoStop}%
\bibitem [{\citenamefont {Mishra}\ \emph {et~al.}(2018)\citenamefont {Mishra},
  \citenamefont {Saha}, \citenamefont {Vigneshwaran}, \citenamefont {Pal},
  \citenamefont {Kapitaniak},\ and\ \citenamefont {Dana}}]{mishra2018dragon}%
  \BibitemOpen
  \bibfield  {author} {\bibinfo {author} {\bibfnamefont {A.}~\bibnamefont
  {Mishra}}, \bibinfo {author} {\bibfnamefont {S.}~\bibnamefont {Saha}},
  \bibinfo {author} {\bibfnamefont {M.}~\bibnamefont {Vigneshwaran}}, \bibinfo
  {author} {\bibfnamefont {P.}~\bibnamefont {Pal}}, \bibinfo {author}
  {\bibfnamefont {T.}~\bibnamefont {Kapitaniak}}, \ and\ \bibinfo {author}
  {\bibfnamefont {S.~K.}\ \bibnamefont {Dana}},\ }\bibfield  {title} {\enquote
  {\bibinfo {title} {Dragon-king-like extreme events in coupled bursting
  neurons},}\ }\href@noop {} {\bibfield  {journal} {\bibinfo  {journal}
  {Physical Review E}\ }\textbf {\bibinfo {volume} {97}},\ \bibinfo {pages}
  {062311} (\bibinfo {year} {2018})}\BibitemShut {NoStop}%
\bibitem [{\citenamefont {Chowdhury}\ \emph {et~al.}(2022)\citenamefont
  {Chowdhury}, \citenamefont {Ray}, \citenamefont {Dana},\ and\ \citenamefont
  {Ghosh}}]{chowdhury2022extreme}%
  \BibitemOpen
  \bibfield  {author} {\bibinfo {author} {\bibfnamefont {S.~N.}\ \bibnamefont
  {Chowdhury}}, \bibinfo {author} {\bibfnamefont {A.}~\bibnamefont {Ray}},
  \bibinfo {author} {\bibfnamefont {S.~K.}\ \bibnamefont {Dana}}, \ and\
  \bibinfo {author} {\bibfnamefont {D.}~\bibnamefont {Ghosh}},\ }\bibfield
  {title} {\enquote {\bibinfo {title} {Extreme events in dynamical systems and
  random walkers: A review},}\ }\href@noop {} {\bibfield  {journal} {\bibinfo
  {journal} {Physics Reports}\ }\textbf {\bibinfo {volume} {966}},\ \bibinfo
  {pages} {1--52} (\bibinfo {year} {2022})}\BibitemShut {NoStop}%
\bibitem [{\citenamefont {Balcerzak}, \citenamefont {Pikunov},\ and\
  \citenamefont {Dabrowski}(2018)}]{balcerzak2018fastest}%
  \BibitemOpen
  \bibfield  {author} {\bibinfo {author} {\bibfnamefont {M.}~\bibnamefont
  {Balcerzak}}, \bibinfo {author} {\bibfnamefont {D.}~\bibnamefont {Pikunov}},
  \ and\ \bibinfo {author} {\bibfnamefont {A.}~\bibnamefont {Dabrowski}},\
  }\bibfield  {title} {\enquote {\bibinfo {title} {The fastest, simplified
  method of lyapunov exponents spectrum estimation for continuous-time
  dynamical systems},}\ }\href@noop {} {\bibfield  {journal} {\bibinfo
  {journal} {Nonlinear Dynamics}\ }\textbf {\bibinfo {volume} {94}},\ \bibinfo
  {pages} {3053--3065} (\bibinfo {year} {2018})}\BibitemShut {NoStop}%
\bibitem [{\citenamefont {Cavalcante}\ \emph {et~al.}(2013)\citenamefont
  {Cavalcante}, \citenamefont {Ori{\'a}}, \citenamefont {Sornette},
  \citenamefont {Ott},\ and\ \citenamefont
  {Gauthier}}]{cavalcante2013predictability}%
  \BibitemOpen
  \bibfield  {author} {\bibinfo {author} {\bibfnamefont {H.~L. d.~S.}\
  \bibnamefont {Cavalcante}}, \bibinfo {author} {\bibfnamefont
  {M.}~\bibnamefont {Ori{\'a}}}, \bibinfo {author} {\bibfnamefont
  {D.}~\bibnamefont {Sornette}}, \bibinfo {author} {\bibfnamefont
  {E.}~\bibnamefont {Ott}}, \ and\ \bibinfo {author} {\bibfnamefont {D.~J.}\
  \bibnamefont {Gauthier}},\ }\bibfield  {title} {\enquote {\bibinfo {title}
  {Predictability and suppression of extreme events in a chaotic system},}\
  }\href@noop {} {\bibfield  {journal} {\bibinfo  {journal} {Physical Review
  Letters}\ }\textbf {\bibinfo {volume} {111}},\ \bibinfo {pages} {198701}
  (\bibinfo {year} {2013})}\BibitemShut {NoStop}%
\end{thebibliography}%
\end{document}